\documentclass[final,5p,times,twocolumn]{elsarticle} 

\usepackage{amssymb}
\usepackage{lineno}

\journal{Nuclear Physics B}

\begin{document}

\begin{frontmatter}



\title{Overview of ATLAS forward proton detectors for LHC Run 3 and plans for the HL-LHC}


\author{Maciej Trzebi\'nski, on behalf of ATLAS Forward Detectors}

\affiliation{organization={Institute of Nuclear Physics Polish Academy of Sciences},%Department and Organization
            addressline={Radzikowskiego 152}, 
            city={Krakow},
            postcode={31-342}, 
            state={},
            country={Poland}}

\begin{abstract}
The status of the ATLAS Roman Pot detectors (AFP and ALFA) for LHC Run 3 after all refurbishments and improvements done during Long Shutdown~2 is discussed.
%Based on analysis of Run 2 data, the expected performance of the Tracking and Time-of-Flight Detectors, the electronics, the trigger, and the readout and detector control and data quality monitoring are described. Finally, 
%The physics interest and the most recent studies of beam optics and detector options for participation at the HL-LHC are shown.
\end{abstract}




\end{frontmatter}

%\linenumbers

\section{ATLAS Forward Proton Detectors}
Diffractive processes are an important part of the physics programme at hadron colliders. This is also true for ATLAS \cite{ATLAS}, where a large community works on both phenomenological and experimental aspects of diffraction. ATLAS Forward Proton detectors (AFP) \cite{AFP} aim to measure events in which one or both protons remain intact after the interaction \cite{LHC_forward_physics}. Since these protons are scattered at very small angles, dedicated detectors must be located far away from the interaction point and close to the proton beam. This results in proton trajectories being influenced by the LHC components: magnets and collimators \cite{Trzebinski_optics}. Since settings of such components change over time, detectors must have a possibility to move wrt. the accelerator beam pipe. This is realised using the Roman Pot (RP) technology.

AFP consists of four RP stations \cite{AFP}, two on each side of ATLAS.
%Detectors located on the ATLAS C side are inserted into the beam 1 whereas the ones on the A side into the beam 2.
Stations located 204 m from ATLAS collision point are called NEAR whereas those at 217 m are named FAR. All stations contain 3D edgeless Silicon Trackers (SiT). FAR stations are also equipped with a Time-of-Flight system (ToF).

\section{Silicon Tracker}
The purpose of the SiT is to provide a precise reconstruction of the proton trajectory necessary to unfold its kinematics \cite{unfolding}. Each station houses four detector planes, each consisting of a 230 $\mu$m thick silicon sensor. Sensor is a matrix of $336 \times 80$ pixels, each having size of $50 \times 250$ $\mu$m$^2$. Sensors are coupled to the FE-I4b chip \cite{fei4b}, radiation-hard (tested for $>250$ Mrad) and produced with the ``edgeless'' technology with a dead edge from beam side of only about 100 $\mu$m. Detectors are tilted by 14 degrees wrt. beam direction. The expected reconstruction resolution is 6 $\mu$m in $x$ and 30 $\mu$m in $y$ \cite{sit}. SiT can provide the trigger signal with a dead-time of about $400$ ns.

During Run 2 data-taking SiT detectors showed very good efficiency. As discussed in Ref. \cite{sit_perf_proc}, NEAR stations had an overall efficiency over 98\% whereas FAR stations performed slightly worse, 95\% -- 98\%. A possible explanation is the radiation degradation of the silicon tracker, as the FAR stations are inserted slightly closer (by about 1 mm) to the beam and are more exposed to the beam halo. In addition, FAR station efficiency is affected by the showers created by interactions with detector material in the NEAR stations.

\section{Time-of-Flight Detectors}
The purpose of the ToF is to reduce the combinatorial background coming from the pile-up, denoted as $\mu$ -- multiple, independent proton-proton collisions during a single bunch crossing. The idea is to measure difference in the time of flight of scattered protons on both sides obtaining a ``proton vertex'' and compare it to the vertex position reconstructed by ATLAS. AFP ToF uses 16 L-shaped quartz bars (LQBars) to produce Cherenkov light and guide it into a Micro-Channel Plate Photo Multiplier (MCP-PMT). During LHC Run 2 the measured time resolution was $35 \pm 6$ ps and $37 \pm 6$ ps per train, depending on side \cite{tof}. This translates to a spatial resolution of $5.2 \pm 0.9$ mm for the vertex $z$-position. AFP ToF is equipped with radiation-hard readout and provides a trigger signal with a dead-time smaller than 25 ns ($<1$ bunch-crossing). Unfortunately, as described in Ref. \cite{tof}, during the Run 2 data-taking ToF system suffered from very low efficiency (a few percent).

%\begin{figure}[!htbp]
%  \centering
%\includegraphics[width=0.35\textwidth]{AFP_ToF_perf.png}
%\caption{The distributions of $z_{ATLAS} - z_{ToF}$ measured in events with ToF signals on both sides of the interaction region. Detailed discussion is in Ref. \cite{tof}.}
%\label{AFP_tof_perf}
%\end{figure}

\section{Readiness for Run 3 Data-taking}
During Run 2 data-taking AFP collected 32 fb$^{-1}$ during high-$\mu$ runs and took part in few low-$\mu$ runs. Afterward, AFP underwent a significant refurbishment. In order to address the major issue of ToF system -- the very low efficiency and repeated failures of MCP-PMTs in vacuum -- the new design of the detector flange (Out-of-Vacuum solution) was made. In addition, a new design of the MCP-PMT back-end electronics was developed. In Run 3 ToF system is also equipped with a set of new, glue-less LQBars. The AFP tracker system was equipped with newly produced SiT modules. In addition, new heat exchangers were installed to improve cooling capabilities. Finally, developments of a new trigger module and picosecond Time to Digital Converter (TDC) readout are ongoing.

\section{High Luminosity LHC Consideration}
At the HL-LHC high pile-up environment the key focus is on photon-induced processes and Beyond Standard Model (BSM) searches. The presence of RPs during Run 4 may improve the measure/search capabilities of ``central'' detectors. The discussion about HL-LHC physics case with use of RPs is held in \cite{CTPPS} and references within \cite{ATLAS_HLLHC}. Below only an idea is described.

Processes which are particularly interesting to be studied are $pp \to pXp$, where $p$ denotes a proton staying intact and $X$ denotes a ``central'' system, \textit{e.g.} photon-induced WW ($\gamma\gamma \to WW$) production, axion-like particles $\gamma\gamma \to a \to \gamma\gamma$, supersymmetry and dark matter particles ($\gamma\gamma \to \tilde{l}\ \tilde{l}$). The information from the forward protons, combined with that of the central detector where particles are produced out of the energy lost by the protons, allows full kinematic reconstruction of the events using the difference and sum of the energy lost by the two protons, respectively.

The presence of pileup makes things more complicated, because the association between the central and forward systems is not obvious, and the two protons tagged by the forward detectors are likely originating from different collisions. For this reason, and especially in the extreme pile-up scenarios foreseen at the HL-LHC, it is essential that the position measurements from the pixel detectors is complemented by a high-resolution ToF detectors.

\begin{figure}[!htbp]
  \centering
\includegraphics[width=0.4\textwidth]{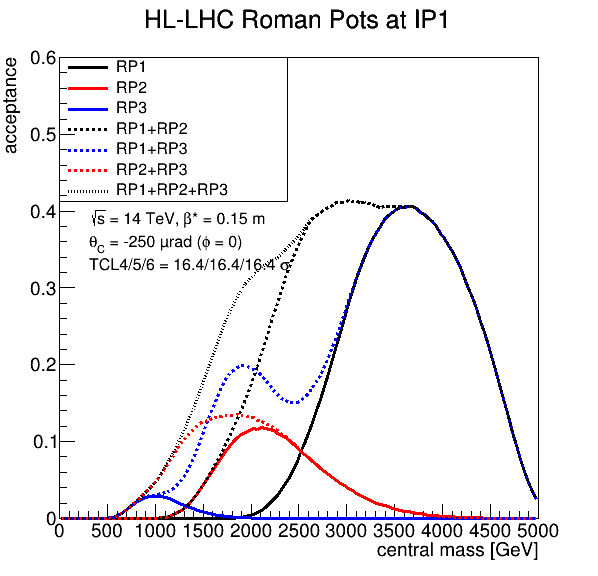}
\caption{Studies for HL-LHC: mass acceptance for considered sets of stations around ATLAS collision point. See text for details.}
\label{HLLHC}
\end{figure}

As there are significant constraints on using RPs due to the HL-LHC layout, a discussion of project feasibility is ongoing within the ATLAS community\footnote{At the moment of writing these proceedings, the decision was taken to not have RPs around ATLAS during Run 4, with an opened possibility for Run 5 and beyond.}. According to the newest available HL-LHC machine layout only a few locations are possible for the RP installation: 195.5 m (RP1A), 198.0 m (RP1B), 217.0 m (RP2A), 219.5 m (RP2B), 234.0 m (RP3A), 237.0 m (RP3B) and 245.0 m (RP3C). Depending on the number of stations and their location, the mass acceptance was estimated -- see Fig. \ref{HLLHC}. The legend should be interpreted as:
\begin{itemize}
  \item ``RP\textbf{X}'' means combination two stations RP\textbf{X}A and RP\textbf{X}B on both sides of IP when proton is tagged in all of them;
  \item ``RP\textbf{X}+RP\textbf{Y}'' means tagged proton in [(RP\textbf{X}A and RP\textbf{X}B on side A) or (RP\textbf{Y}A and RP\textbf{Y}B on side A)] and [(RP\textbf{X}A and RP\textbf{X}B on side C) or (RP\textbf{Y}A and RP\textbf{Y}B on side C)].
\end{itemize}

\vspace*{-0.2cm}
\section{Summary}
During Long Shutdown 2, AFP underwent hardware upgrades before being deployed in the LHC tunnel. This should allow efficient data-taking with a focus on a diffractive and minimum-bias studies as well as BSM searches during Run 3. The presence of Roman Pots in High-Luminosity LHC is being discussed within ATLAS as a variety of physics studies would benefit from the forward proton tagging. This would require the development of a rather complex integration design with the new HL-LHC layout.

\section*{Acknowledgements}
The work of MT was partially supported by Polish National Science Centre (project no. UMO-2019/34/E/ST2/00393).\\
Copyright 2022 CERN for the benefit of the ATLAS Collaboration. Reproduction of this article or parts of it is allowed as specified in the CC-BY-4.0 license.




\begin{thebibliography}{0}
{

\bibitem{ATLAS} ATLAS Collaboration, \textit{The ATLAS Experiment at the CERN Large Hadron Collider}, JINST \textbf{3} (2008) S08003

\bibitem{AFP} ATLAS Collaboration, \textit{Technical Design Report for the ATLAS Forward Proton Detector}, CERN-LHCC-2015-009, ATLAS-TDR-024

\bibitem{LHC_forward_physics} K. Akiba et al., \textit{LHC Forward Physics}, J. Phys. G: Nucl. Part. Phys. \textbf{43} (2016) 110201, arXiv:1611.05079 [hep-ph], https://doi.org/10.48550/arXiv.1611.05079
%
\bibitem{Trzebinski_optics} M. Trzebinski, \textit{Machine Optics Studies for the LHC Measurements}, in proceedings of XXXIV-th IEEE-SPIE Joint Symposium Wilga 2014, SPIE 0277-786X, vol. 9290 (2014) 26, arXiv:1408.1836 [physics.acc-ph], https://doi.org/10.1117/12.2074647
%
\bibitem{unfolding} M. Trzebinski, R. Staszewski, J. Chwastowski, \textit{LHC High Beta* Runs: Transport and Unfolding Methods}, ISRN High Energy Physics \textbf{2012} (2012) 491460,  	arXiv:1107.2064 [physics.ins-det], https://doi.org/10.5402/2012/491460
%
\bibitem{fei4b} M. Backhaus, \textit{Characterization of the FE-I4B pixel readout chip production run for the ATLAS Insertable B-layer upgrade}, JINST \textbf{8} (2013) C03013.
%
\bibitem{sit} J. Lange et al., \textit{Beam tests of an integrated prototype of the ATLAS Forward Proton detector}, JINST \textbf{11} (2016) P09005, arXiv:1608.01485 [physics.ins-det], https://doi.org/10.1088/1748-0221/11/09/P09005
%
\bibitem{sit_perf_proc} L. Alcerro, G. K. Krintiras, C. and Royon, \textit{Proceedings of the Low-$x$ 2021 International Workshop}, arXiv:2206.11624 [hep-ph]
%
\bibitem{tof} ATLAS Collaboration, \textit{Performance of the ATLAS Forward Proton Time-of-Flight Detector in 2017}, ATL-FWD-PUB-2021-002, https://cds.cern.ch/record/2749821
%
\bibitem{CTPPS} CMS Collaboration, \textit{The CMS Precision Proton Spectrometer at the HL-LHC -- Expression of Interest}, CERN-CMS-NOTE-2020-008, arXiv:2103.02752 [physics.ins-det],  	
https://doi.org/10.48550/arXiv.2103.02752
%
\bibitem{ATLAS_HLLHC} J. Liu, \textit{Future opportunities: AFP-HL-LHC}, presentation during \textit{Photon induced processes} workshop, Durham 3-4/11/2022, https://conference.ippp.dur.ac.uk/event/1143/
%\bibitem{ATLAS_public} ATLAS Collaboration, \textit{Public Forward Detector Plots for Collision Data}, https://twiki.cern.ch/twiki/bin/view/AtlasPublic/ForwardDetPublicResults
%

}
\end{thebibliography}


\end{document}